\documentclass[letterpaper,onecolumn]{IEEEtranBB}   

\usepackage{xspace}
\usepackage{mathrsfs}
\usepackage{amsopn}

\newcommand{\Sc}{\mathcal{S}}

\newcommand{\Xc}{\mathcal{X}}
\newcommand{\Yc}{\mathcal{Y}}

\newcommand{\Rr}{\mathscr{R}}

\def\eps{\epsilon}

\let\P\relax
\DeclareMathOperator\P{P}

\def\textiid{i.i.d.\@\xspace}
\newcommand\iid{\ifmmode\text{ i.i.d. } \else \textiid \fi}

\newcommand{\Real}{\mathbb{R}}

\usepackage[T1]{fontenc}
\usepackage{graphicx}
	\graphicspath{{figs/}{matlab/}}   
\usepackage[cmex10]{amsmath}
\usepackage{amssymb, amsthm, gensymb}  
\usepackage{booktabs}  
	\renewcommand{\tabcolsep}{4mm}  
\usepackage{cellspace}
\usepackage{float}
\usepackage{flafter}  
\usepackage{accents}
\usepackage{cases}  
\usepackage[usenames,dvipsnames]{color}

\usepackage{calc}

\newcommand{\executeiffilenewer}[3]{%
\ifnum\pdfstrcmp{\pdffilemoddate{#1}}%
{\pdffilemoddate{#2}}>0%
{\immediate\write18{#3}}\fi%
}
\newcommand{\includesvg}[1]{%
\executeiffilenewer{#1.svg}{#1.pdf}%
{inkscape -z -D --file=#1.svg %
--export-pdf=#1.pdf --export-latex}%
\ifx\svgscale\undefined%
	\newcommand{\svgscale}{\defaultsvgscale}
\fi%
\input{#1.pdf_tex}%
}

	\newcommand{\defaultsvgscale}{1.0}  

\usepackage[tight,small,bf]{subfigure}   

\usepackage{microtype}

\usepackage{caption}
\DeclareCaptionLabelSeparator{periodquad}{. \quad}
\ifCLASSOPTIONonecolumn
	\ifCLASSOPTIONtrimmargin
		\usepackage{anysize}
		\papersize{9.7in}{5.9in}  
		\marginsize{0.2in}{0.2in}{0.1in}{0.1in} 
	\else
		\usepackage{anysize}
		\marginsize{1.5in}{1.5in}{0.75in}{0.75in}  
	\fi
\fi

\usepackage[colorlinks=true,linkcolor=black,anchorcolor=black,citecolor=black,filecolor=black,menucolor=black,urlcolor=black]{hyperref}  
	
\usepackage{cite}  

\renewcommand{\eps}{\varepsilon}

\definecolor{unemphColor}{gray}{0.5}  

\newcommand{\CH}[1]{\overline{#1}}
\newcommand{\natSetVar}[2]{\{#1\hspace{-0.25em}:\hspace{-0.25em}#2\}}  
\newcommand{\natSet}[1]{\natSetVar{1}{#1}}  

\newcommand{\dhat}[1]{\ring{#1}}
\newcommand{\Typ}{\mathcal T_\varepsilon^{(n)}}
\newcommand{\Typprime}{\mathcal T_{\varepsilon'}^{(n)}}
\newcommand{\Er}{\mathcal E}
\newcommand{\cond}{\,|\,}
\newcommand{\condbig}{\bigm|}

\theoremstyle{definition}  
\newtheorem{thm}{Theorem}

\theoremstyle{remark}

\newcommand{\ReduceA}{\mathop{\mathrm{FM}}}
\newcommand{\Es}{\mathcal E_\text{eq}}
\newcommand{\Lc}{\mathcal{L}}
\newcommand{\Sr}{\mathscr{S}}

\title{An Achievable Rate Region for \\Three-Pair Interference Channels \\with Noise}   
\IEEEoverridecommandlockouts
\author{Bernd Bandemer\\
Information Theory and Applications Center,  \\
University of California, San Diego \\
La Jolla, CA 92093, USA \\
Email: bandemer@ucsd.edu
\thanks{\hrule \vspace{2mm} \noindent This research was supported in part by the Korea Communications Commission under the R\&D program KCA-2012-11-921-04-001 (ETRI).}
\thanks{A slightly abbreviated version of this work was accepted for presentation at ISIT 2012, Boston, MA. This version contains additional material in Section~\ref{sec:HK} (Han--Kobayashi inner bound).
}
}

\hypersetup{
	pdfauthor = {Bernd Bandemer},
	pdftitle = {An Achievable Rate Region for Three-Pair Interference Channels with Noise},
	pdfsubject = {Information theory},
	pdfkeywords = {},
	pdfcreator = {LaTeX}
}

\begin{document}
\maketitle


\begin{abstract} 
An achievable rate region for certain noisy three-user-pair interference channels is proposed. The channel class under consideration generalizes the three-pair deterministic interference channel (3-DIC) in the same way as the Telatar--Tse noisy two-pair interference channel generalizes the El\;Gamal--Costa  injective channel. Specifically, arbitrary noise is introduced that acts on the combined interference signal before it affects the desired signal. This class of channels includes the Gaussian case.

The rate region includes the best-known inner bound on the 3-DIC capacity region, dominates treating interference as noise, and subsumes the Han--Kobayashi region for the two-pair case.
\end{abstract}

\vspace{2mm}
\section{Introduction}
The interference channel is one of the canonical models in network information theory, and has withstood all attempts to solve it in general. In recent years, significant progress has been made for the case with two sender--receiver pairs. 
The best known achievable rate region is achieved by the Han--Kobayashi coding scheme~\cite{HanKobayashi81}, for which a compact formulation was given in~\cite{ChongMotani08}. 
Much less is known in the case with more than two user pairs. Major lines of work exist in the areas of interference alignment~\cite{MaddahAli2008,CadambeJafar2008}, and deterministic models as pioneered in~\cite{Avestimehr07, Avestimehr2011}. The key idea in the latter is to first investigate a simplified interference channel that does not contain noise, and then proceed to transfer the insight to more practically relevant channels with noise.

\begin{figure}
	\centering
	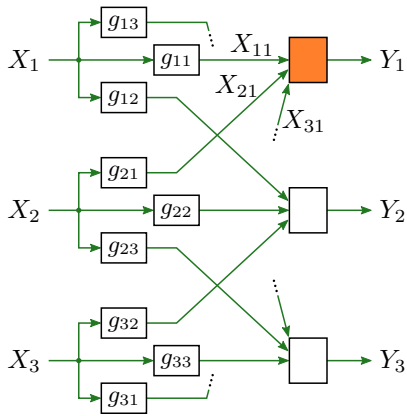
	\caption{Three-pair interference channel under consideration.}
	\label{fig:f_3dic}
\end{figure}

\begin{figure}
	\centering
	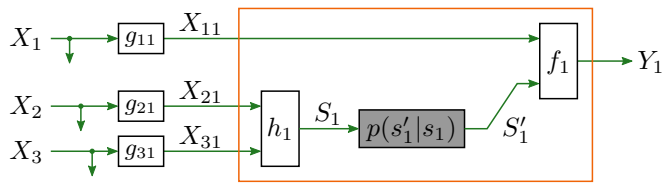
	\caption{Channel under consideration, as seen by the first receiver.}
	\label{fig:f_3nic_1stReceiver}
\end{figure}

In this paper, we apply this idea to the three-user-pair deterministic interference channel (3-DIC) first introduced in~\cite{Bandemer:2010:InterferenceDecoding}. 
We consider the noisy version of the 3-DIC depicted in Figures~\ref{fig:f_3dic} and~\ref{fig:f_3nic_1stReceiver}.
The channel consists of three sender--receiver alphabet pairs $(\Xc_l,\Yc_l)$, loss functions $g_{lk}$ that model the links between each sender and receiver, and a conditional probability mass function (pmf) at each receiver that maps the three impinging signals into the receiver observation $Y_l$, for indices $k,l \in \natSet 3$.
The pmfs have the special structure depicted in Figure~\ref{fig:f_3nic_1stReceiver} for the first receiver. They consist of two deterministic stages, namely an interference combining function $h_l$ and a receiver function $f_l$. We assume that the functions $h_l$ and $f_l$ are injective in each argument, 
that is, they become one-to-one when either one of their arguments is fixed. For example, for $Y_1 = f_1(X_{11}, S_1')$, this assumption is equivalent to $H(X_{11})=H(Y_1 \cond S_1')$ and $H(S_1')=H(Y_1 \cond X_{11})$ for every pmf $p(x_{11}, s_1')$. An example of a function that is injective in each argument (but not injective) is regular addition.
Deviating from the deterministic nature of the 3-DIC, we introduce noise between the two combining stages. It acts on the combined interference signal $S_l$ and is characterized by the discrete memoryless channel $p(s_l'|s_l)$. (Note that setting $S'_l=S_l$ for all $l$ recovers the 3-DIC setting.)

Each sender $l$ wishes to convey an independent message $M_l$ at rate $R_l$ to its corresponding receiver. We define a $(2^{nR_1},2^{nR_2},2^{nR_3},n)$ code, probability of error, achievability of a rate triple $(R_1,R_2,R_3)$, and the capacity region in the standard way (see~\cite{ElGamalKim}). The capacity region is not known, but in this paper, we make progress towards characterizing it.

The channel model under consideration is interesting since it contains the Gaussian interference channel as a special case. Characterizing its capacity region thus has immediate consequences for practical wireless communications systems where simultaneous transmissions use the same radio spectrum. Moreover, we follow along the footsteps of previous work in the case with two user pairs. The capacity region of the two-pair version of 3-DIC was found in~\cite{ElGamalCosta82}. A modification of this channel in which noise affects the interfering signal was studied in~\cite{TelatarTse07}. The authors establish inner and outer bounds to the capacity region which differ only by a constant gap, akin to the Gaussian case studied in~\cite{EtkinTse08}. 

Thus we are motivated to generalize the results for the 3-DIC to its own noisy cousin.
Let us briefly review the 3-DIC results. 
The best known achievable rate region is given in~\cite{Bandemer:2011:3DicMarton}. The underlying scheme combines insights from the transmitter-centric view of communication with disturbance constraints~\cite{Bandemer:2011:DisturbanceJournal} and the receiver-centric view of interference decoding~\cite{Bandemer:2010:InterferenceDecodingJournal}.  All results of \cite{Bandemer:2010:InterferenceDecodingJournal,Bandemer:2011:DisturbanceJournal,Bandemer:2011:3DicMarton} are also contained and discussed in detail in~\cite{Bandemer:2011:PhdThesis}.

We extend the achievable rate region in~\cite{Bandemer:2011:3DicMarton} to the channel under consideration. It turns out that the key properties of 3-DIC are preserved and allow us to apply the same coding scheme, which consists of rate splitting, Marton coding, and superposition coding. The analysis of the probability of error is more involved than in the deterministic case due to the noise.
The resulting inner bound to the capacity region includes all previous results for the 3-DIC. It simplifies to the Han--Kobayashi inner bound when one of the three user pairs is not present. 
Finally, unlike the interference decoding inner bound for the 3-DIC with noisy observations~\cite{Bandemer:2010:InterferenceDecodingJournal}, the present bound is larger than the one that results from using point-to-point random codes and treating interference as noise.

\vspace{3mm} 
\section{Achievable rate region} \label{sec:result}
In order to state the inner bound to the capacity region of the channel under consideration, we need the following definitions.
Fix a joint pmf for $(Q,U_1,X_1,U_2,X_2,U_3,X_3)$ of the form $$p = p(q)\,p(u_1,x_1|q)\,p(u_2,x_2|q)\,p(u_3,x_3|q).$$ 
Define the rate region $\Rr_1(p) \subset \Real_+^{18}$ to consist of the rate tuples 
\begin{align}
(  R_{10}, R_{11},R_{12},R_{13},\tilde R_{12},\tilde R_{13}, 
R_{20}, R_{22},R_{23},R_{21},\tilde R_{23},\tilde R_{21}, 
R_{30}, R_{33},R_{31},R_{32},\tilde R_{31},\tilde R_{32}  )
\label{eq:18dimRateVector}
\end{align}
that satisfy
{\allowdisplaybreaks
\begin{align}
	\tilde R_{12} - R_{12} + \tilde R_{13} - R_{13} &\geq I(X_{12}; X_{13} \cond U_1, Q), \label{eq:resultIneq1} \\
	\tilde R_{12} - R_{12} + (\tilde R_{13} - R_{13})/2 &\leq I(X_{12}; X_{13} \cond U_1, Q), \label{eq:resultMaxExcess1} \\
	(\tilde R_{12} - R_{12})/2 + \tilde R_{13} - R_{13} &\leq I(X_{12}; X_{13} \cond U_1, Q), \label{eq:resultMaxExcess2} \\
	\tilde R_{12} & \geq R_{12}, \\
	\tilde R_{13} & \geq R_{13} \label{eq:resultIneq2},
\end{align}
}%
and
\begin{align}
	& \forall i \in \natSet 5, j \in \natSet 3, k \in \natSet 3:  \exists  j', k':  \notag \\
	& \hspace{20mm}r_{1i} + r_{21j'} + r_{31k'} + I(S_1;S_1'\cond c_{21j'}, c_{31k'}, Q)  \notag \\
	&\hspace{20mm}\qquad \qquad \leq I(X_1,X_2,X_3; Y_1 \cond c_{1i}, c_{21j}, c_{31k}, Q) + t_{1i}.
	\label{eq:indexedIneq}
\end{align}
In~\eqref{eq:indexedIneq}, symbols like $r_{1i}$, $c_{1i}$, and $t_{1i}$ are placeholders for the terms specified in Tables~\ref{tab:shorthand1}, \ref{tab:shorthand2}, and~\ref{tab:shorthand3}, respectively. 
For example, for $i=3$, $j=3$, and $k=2$, condition~\eqref{eq:indexedIneq} becomes
\begin{align}
	& \tilde R_{13} + R_{11} + \min \bigl\{I(S_1;S_1'\cond U_3, Q), \notag \\
	& \hspace{24mm} R_{20} + I(S_1;S_1'\cond U_2, U_3, Q), \notag \\
	& \hspace{24mm} R_{20} + \tilde R_{21} + I(S_1;S_1'\cond X_2, U_3, Q), \notag \\
	& \hspace{24mm} \tilde R_{31} + I(S_1;S_1'\cond X_3, Q), \notag \\
	& \hspace{24mm} R_{20} + \tilde R_{31} + I(S_1;S_1'\cond U_2, X_3, Q), \notag \\
	& \hspace{24mm} R_{20} + \tilde R_{21} + \tilde R_{31} 
	\bigr\}   
	\quad \leq \quad I(X_1,X_2,X_3; Y_1 \cond U_1,X_{12}, U_3, Q) \notag \\
	& \hspace{80.5mm} + I(X_{12}; X_{13} \cond U_1, Q). 
	\label{eq:332_expanded}
\end{align}


\begin{table}
	\vspace{3mm}
	\centering
	\small
	\begin{tabular}{llll}
		\toprule
		$i$ & $r_{1i}$ & $c_{1i}$ & $t_{1i}$ \\  \midrule
		1 & $R_{11}$ & $\{ U_1, X_{12}, X_{13} \}$ & $0$ \\
		2 & $\tilde R_{12} + R_{11}$ & $\{ U_1, X_{13} \}$ & $I_1$ \\
		3 & $\tilde R_{13} + R_{11}$ & $\{ U_1, X_{12} \}$ & $I_1$ \\
		4 & $\tilde R_{12} + \tilde R_{13} + R_{11}$ 
			& $\{ U_1\}$ & $I_1$ \\
		5 & $R_{10} + \tilde R_{12} + \tilde R_{13} + R_{11}$ 
			& $\emptyset$ & $I_1$ \\ \bottomrule
	\end{tabular}
	\caption{Shorthand notation for terms related to transmitter $1$. The term $I_1$ stands for $I(X_{12}; X_{13} \cond U_1, Q)$.}
	\label{tab:shorthand1}
\end{table}

\begin{table}
	\vspace{3mm}
	\centering
	\small
	\begin{tabular}{lllll}
		\toprule
		$j$ & $c_{21j}$ & $j'$ & $r_{21j'}$ & $c_{21j'}$ \\ \midrule
		1 & $\{X_2\}$ & 1 & $0$ & $\{X_2\}$ \\\addlinespace[2mm]
		2 & $\{U_2\}$ & 1 & $0$ & $\{U_2\}$ \\
		  &           & 2 & $\tilde R_{21}$ & $\{X_2\}$ \\\addlinespace[2mm]
		3 &$\emptyset$& 1 & $0$  & $\emptyset$ \\
		  &           & 2 & $R_{20}$ \phantom{$\tilde R$}& $\{U_2\}$ \\
		  &           & 3 & $R_{20}+\tilde R_{21}$ & $\{X_2\}$ 	\\ \bottomrule
	\end{tabular}
	\caption{Shorthand notation for terms related to transmitter $2$.}
	\label{tab:shorthand2}
\end{table}

\begin{table}
	\vspace{3mm}
	\centering
	\small
	\begin{tabular}{lllll}
		\toprule
		$k$ & $c_{31k}$ & $k'$ & $r_{31k'}$ & $c_{31k'}$ \\ \midrule
		1 & $\{X_3\}$ & 1 & $0$ & $\{X_3\}$ \\\addlinespace[2mm]
		2 & $\{U_3\}$ & 1 & $0$ & $\{U_3\}$ \\
		  &           & 2 & $\tilde R_{31}$ & $\{X_3\}$ \\\addlinespace[2mm]
		3 &$\emptyset$& 1 & $0$  & $\emptyset$ \\
		  &           & 2 & $R_{30}$ \phantom{$\tilde R$}& $\{U_3\}$ \\
		  &           & 3 & $R_{30}+\tilde R_{31}$ & $\{X_3\}$ 	\\ \bottomrule
	\end{tabular}
	\caption{Shorthand notation for terms related to transmitter $3$.}
	\label{tab:shorthand3}
\end{table}

Similarly, define the regions $\Rr_2(p)$ and $\Rr_3(p)$ by making the subscript replacements $1\mapsto 2 \mapsto 3 \mapsto 1$  and $1\mapsto 3 \mapsto 2 \mapsto 1$ in the definition of $\Rr_1(p)$, respectively. 

Define a Fourier--Motzkin elimination operator $\ReduceA$ that maps a convex $18$-dimensional set of rate vectors of the form~\eqref{eq:18dimRateVector} to a $3$-dimensional region by letting $R_l = \sum_{\nu=0}^3 R_{l\nu}$, for $l\in \natSet 3$, and  projecting on the coordinates $(R_1,R_2,R_3)$. Let $\CH{\Sr}$ denote the convex hull of $\Sr$. 

We are now ready to state the main result as follows.

\begin{thm}[Achievable rate region] \label{thm:ach}  The region
\begin{align*}
\Rr=\ReduceA\left\{ \CH{ \bigcup_{p} \bigl( \Rr_1(p) \cap \Rr_2(p) \cap \Rr_3(p) \bigr) } \right\}, 
\end{align*}
where $p=p(q)\,p(u_1,x_1|q)\,p(u_2,x_2|q)\,p(u_3,x_3|q)$, is achievable in the interference channel under consideration.
\end{thm}

The regions $\Rr_1(p)$, $\Rr_2(p)$, and $\Rr_3(p)$ in the theorem represent decodability conditions at the first, second, and third receiver, correspondingly. The regions and their intersection are generally nonconvex. By the time-sharing argument, we are allowed to convexify, as shown in the theorem. This convex hull operation is nontrivial even for a fixed pmf $p$, and therefore, it is not automatically achieved by including $Q$. Due to the explicit convex hull operation, the Fourier--Motzkin reduction $\ReduceA$ cannot be evaluated symbolically.

\subsection{Coding scheme}
We outline the coding scheme that attains the inner bound of Theorem~\ref{thm:ach}. To simplify the notation, we omit the time-sharing auxiliary random variable $Q$ throughout this section.

\vspace{2mm} \noindent \emph{Codebook generation: } 
The transmitter codebooks are generated as in~\cite{Bandemer:2011:3DicMarton}, inspired by communication with disturbance constraints~\cite{Bandemer:2011:DisturbanceJournal}. The intuition is that the interference channel under consideration is sufficiently deterministic in nature such that the results from the deterministic case of communication with disturbance constraints still apply. In particular, disturbance is measured before the combining functions $h_l$, and thus before the noise appears.

Fix a pmf $p(u_1,x_1)\,p(u_2,x_2)\,p(u_3,x_3)$. Consider the first transmitter. Split the rate as $R_1=R_{10}+R_{11}+R_{12}+R_{13}$, and define the auxiliary rates $\tilde R_{12} \geq R_{12}$ and $\tilde R_{13} \geq R_{13}$. Let $\eps'>0$, and define the set  partitions 
\begin{align*}
	\natSet{2^{n\tilde R_{12}}} &= \Lc_{12}(1) \cup \dots \cup \Lc_{12}(2^{nR_{12}}), \\
	\natSet{2^{n\tilde R_{13}}} &= \Lc_{13}(1) \cup \dots \cup \Lc_{13}(2^{nR_{13}}),
\end{align*}
where $\Lc_{12}(\cdot)$ and $\Lc_{13}(\cdot)$ are indexed sets of size $2^{n(\tilde R_{12}-R_{12})}$ and $2^{n(\tilde R_{13}-R_{13})}$, respectively. 
\begin{enumerate}
	\item For each $m_{10} \in \natSet{2^{nR_{10}}}$, generate $u_1^n(m_{10})$ according to $\prod_{i=1}^n p(u_{1i})$.
	\item For each $l_{12}\in \natSet{2^{n\tilde R_{12}}}$, generate $x_{12}^n(m_{10},l_{12})$ according to $\prod_{i=1}^n p(x_{12i}\cond u_{1i}(m_{10}))$. Likewise, for each $l_{13}\in \natSet{2^{n\tilde R_{13}}}$, generate a sequence $x^n_{13}(m_{10}, l_{13})$ according to $\prod_{i=1}^n p(x_{13i}\cond u_{1i}(m_{10}))$.\label{item:3dic_MC_Gen_indep1}
	\item For each triple $ (m_{10},m_{12},m_{13})$, let $\Sc(m_{10},m_{12},m_{13})$ be the set of all pairs $(l_{12},l_{13})$ from the product set $\Lc_{12}(m_{12}) \times \Lc_{13}(m_{13})$ such that $( x^n_{12}(m_{10},l_{12}),$ $ x^n_{13}(m_{10},l_{13}) ) \in \Typprime(X_{12}, Z_{13} \cond u_1^n(m_{10})).$ 
	\item For each $(m_{10}, l_{12}, l_{13})$ and $m_{11} \in \natSet{2^{nR_{11}}} $, generate a sequence
			$x_1^n(m_{10}, l_{12},l_{13},m_{11}) $ according to 
$				\prod_{i=1}^n p ( x_{1i} \cond u_{1i}(m_{10}), x_{12i}(l_{12}), x_{13i}(l_{13}) )
$,			
		if $(l_{12},l_{13}) \in \Sc(m_{10},m_{12},m_{13})$. Otherwise, generate it according to $\mathrm{Unif}(\Xc_1^n)$.
	\label{item:3dic_MC_Gen_superpose}
	\item Draw a random pair uniformly from $\Sc(m_{10},m_{12},m_{13})$ and denote it as $(l_{12}^{(m_{10},m_{12},m_{13})} ,$ $  l_{13}^{(m_{10},m_{12},m_{13})})$. If $\Sc(m_{10},m_{12},m_{13})$ is empty, use $(1,1)$ instead.\label{item:3dic_MC_Gen_Error}		
\end{enumerate}
Codebooks for the second and third transmitter are generated analogously by applying the subscript replacements $1\mapsto 2 \mapsto 3 \mapsto 1$  and $1\mapsto 3 \mapsto 2 \mapsto 1$ in each step of the procedure. 

\vspace{2mm}
\noindent \emph{Encoding: }  To send message $m_1=(m_{10},m_{12},m_{13},m_{11})$, transmit $$x_1^n(m_{10}, l_{12}^{(m_{10},m_{12},m_{13})}, l_{13}^{(m_{10},m_{12},m_{13})}, m_{11}).$$

\vspace{2mm} \noindent \emph{Decoding: } The receivers use simultaneous non-unique decoding~\cite{ElGamalKim}. The first receiver observes $y_1^n$. Define the tuple
\begin{align*}
	&T(m_{10}, m_{12}, m_{13}, m_{11}, m_{20}, l_{21}, m_{30}, l_{31}) \\
	&= \Bigl( 
		u^n_1(m_{10}),
		x^n_{12}(m_{10}, l_{12}^{(m_{10}, m_{12}, m_{13} )}),   \\
	&\qquad
		x^n_{13}(m_{10}, l_{13}^{(m_{10}, m_{12}, m_{13} )}),   \\
	&\qquad   
		x_1^n(m_{10}, l_{12}^{(m_{10}, m_{12}, m_{13} )}, l_{13}^{(m_{10}, m_{12}, m_{13} )}, m_{11} ),  \\
	&\qquad   
		u^n_2(m_{20}), x^n_{21}(m_{20}, l_{21}),
		u^n_3(m_{30}), x^n_{31}(m_{30}, l_{31}), \\
	&\qquad
		s_1^n( m_{20}, l_{21}, m_{30}, l_{31} ), 
		y_1^n 
	\Bigr).
\end{align*}
Declare that $\hat m_1 = (\hat m_{10}, \hat m_{12}, \hat m_{13}, \hat m_{11})$ has been sent if it is the unique message such that 
\begin{align*}
	T(\hat m_{10}, \hat m_{12}, \hat m_{13}, \hat m_{11}, \dhat m_{20}, \dhat l_{21}, \dhat m_{30},\dhat l_{31})
	& \in \Typ(U_1,X_{12},X_{13},X_1,U_2,X_{21},U_3,X_{31},S_1,Y_1)
\end{align*}
for some $\dhat m_{20}, \dhat l_{21}, \dhat m_{30}, \dhat l_{31}$.

\vspace{2mm} \noindent \emph{Analysis of error probability: } 
Each triple $(i,j,k)$ in~\eqref{eq:indexedIneq} corresponds to a certain error event, the probability of which must asymptotically vanish. This can be ensured by any one of several conditions, indexed by $j'$ and $k'$. The details are deferred to the appendix.

\subsection{Discussion}
Different $(j',k')$ in~\eqref{eq:indexedIneq} correspond to various ways of interference signal saturation, as first discussed in~\cite{Bandemer:2010:InterferenceDecodingJournal}. Saturation takes place when the total number of interfering codewords exceeds the number of distinguishable sequences, and thus it is 
not possible to decode the interfering messages.
This is illustrated by the example in~\eqref{eq:332_expanded}. Let us compare the first and the last terms in the $\min$ expression, i.e.,
\begin{align*}
	I(S_1;S_1'\cond U_3,Q)\qquad &\text{and} \qquad R_{20} + \tilde R_{21} + \tilde R_{31}.
\end{align*}
In the noiseless case, 
$S_1'$ equals $S_1$, and the first term becomes $H(S_1\cond U_3,Q)$. In logarithmic scale, this is the size of the set of typical interfering sequences that can appear under the error event in question ($i=3$, $j=3$, $k=2$). Saturation occurs if the interfering rate $R_{20} + \tilde R_{21} + \tilde R_{31}$ exceeds this quantity, and thus increasing the rates does not further increase the set of observed interference sequences. On the other hand, when noise is present, we have
\begin{align*}
	I(S_1;S_1'\cond U_3,Q) &= H(S_1\cond U_3,Q) - H(S_1\cond U_3,S_1',Q) \\
	&\leq  H(S_1\cond U_3,Q),
\end{align*}
which implies that saturation starts to occur at lower rates than in the noiseless case. Loosely speaking, each interfering sequence takes up more of the observed signal space due to channel noise. Along similar lines, the remaining terms in the $\min$ expression in~\eqref{eq:332_expanded} correspond to other modes of (partial) saturation.
Thus, we can interpret the choice of $(j',k')$ in condition~\eqref{eq:indexedIneq} as switching between different regimes of saturation, and treating the saturated sequences as i.i.d.~noise as appropriate. Keep in mind, however, that the entire inner bound is achieved by the same simultaneous non-unique typicality decoder. The distinction of saturation regimes appears only through different ways of analyzing the error probability of the same decoding rule.

It is interesting to note that Theorem~\ref{thm:ach} contains the following three special cases.
\vspace{2mm}
\subsubsection{3-DIC inner bound}
It is not hard to see that Theorem~\ref{thm:ach} subsumes previously known results for the 3-DIC case where $S_l'=S_l$ for all $l$. The inequalities in~\eqref{eq:indexedIneq} for a given triple $(i,j,k)$ are implied by the conditions in~\cite[Corollary 1]{Bandemer:2011:3DicMarton}. In fact, Theorem~\ref{thm:ach} slightly improves upon the results in the deterministic case: For example, comparing the $\min$ terms in~\cite[equation (19)]{Bandemer:2011:3DicMarton} with those in~\eqref{eq:332_expanded} reveals that terms such as $H(X_{21}\cond U_2,Q)+H(X_{31}\cond U_3,Q)$ can be replaced by $H(S_1\cond U_2,U_3,Q)$. The improvement comes from the refined proof technique in the appendix.

\vspace{2mm}
\subsubsection{Point-to-point codes with treating interference as noise} 
Inspecting~\eqref{eq:indexedIneq}, note that
\begin{align*}
	& I(X_1,X_2,X_3; Y_1 \cond c_{1i}, c_{21j}, c_{31k}, Q) \\
	& \qquad - I(S_1;S_1'\cond c_{21j'}, c_{31k'}, Q) \\
	&= H(Y_1\cond c_{1i}, c_{21j}, c_{31k}, Q) \\
	& \qquad - \underbrace{H(Y_1\cond X_1,X_2,X_3, c_{1i}, c_{21j}, c_{31k}, Q)}_{H(S_1'\cond S_1,Q)} \\
	& \qquad - \underbrace{H(S_1'\cond c_{21j'}, c_{31k'}, Q)}_{H(Y_1\cond X_1,c_{1i},c_{21j'}, c_{31k'}, Q)} + \underbrace{H(S_1'\cond S_1, c_{21j'}, c_{31k'}, Q)}_{H(S_1'\cond S_1,Q)} \\
	&= I(X_1, c_{21j'}, c_{31k'} ; Y_1 \cond c_{1i}, c_{21j}, c_{31k}, Q),
\end{align*}
where the last step relies on the Markov chains $c_{21j} - c_{21j'} - Y_1$ and $c_{31k} - c_{31k'} - Y_1$. Therefore, an equivalent way to write condition~\eqref{eq:indexedIneq} is
\begin{align}
	& \forall i \in \natSet 5, j \in \natSet 3, k \in \natSet 3: \exists  j', k':  \notag \\
	& \hspace{20mm}r_{1i} + r_{21j'} + r_{31k'}  \leq I(X_1,c_{21j'}, c_{31k'}; Y_1 \cond c_{1i}, c_{21j}, c_{31k}, Q) + t_{1i}.
	\label{eq:indexedIneq_alt}
\end{align}
This implies that Theorem~\ref{thm:ach} includes the achievable rate region attained by point-to-point random codes and treating interference as noise. To see this, set $U_l=X_l$, let $R_{12}=\tilde R_{12}=R_{13}=\tilde R_{13}=R_{11}=0$ and $R_{10}=R_1$, and likewise for the other transmitters. Only the cases with $i=5$ remain, and we choose $j'=k'=1$. Then condition~\eqref{eq:indexedIneq_alt} is implied by 
\begin{align*}
	\forall j\in \natSet 3, k\in \natSet 3: \quad
	R_1 &\leq I(X_1; Y_1, c_{21j}, c_{31k} \cond Q).
\end{align*}
Among these, $j=k=3$ is most stringent, leading to 
\begin{align*}
	R_1 &\leq I(X_1;Y_1\cond Q),
\end{align*}
which is the rate achievable by using point-to-point (non-layered) random codes and treating interference as noise. The same inclusion does not hold in the case of the interference decoding inner bound for the \emph{3-DIC with noisy observations} studied in~\cite{Bandemer:2010:InterferenceDecodingJournal}, where the noise in the channel acts on $Y_l$ instead of $S_l$. In that case, the saturation effects are exploited without taking the noise into account, which leads to an artificial separation between the channel noise and the combined interference even if the latter is to be treated as noise. In the present case, however, since the noise directly affects the combined interference signal $S_l$, saturation and noise can be treated jointly as discussed above, and the inefficiency of artificially separating them is avoided.

\vspace{2mm} 
\subsubsection{Han--Kobayashi inner bound} \label{sec:HK}
Finally, when one of the three user pairs is not present, say, the third one ($X_{13}=X_{23}=X_3=\emptyset$), Theorem~\ref{thm:ach} recovers the Han--Kobayashi inner bound for the interference channel that consists of the first and second user pair. 

\begin{figure}[h!]
	\centering
	\subfigure[With $\tilde R_{12}=R_{12}$ and $\tilde R_{13}=R_{13}=0$.]{
		\hspace*{-10mm}
		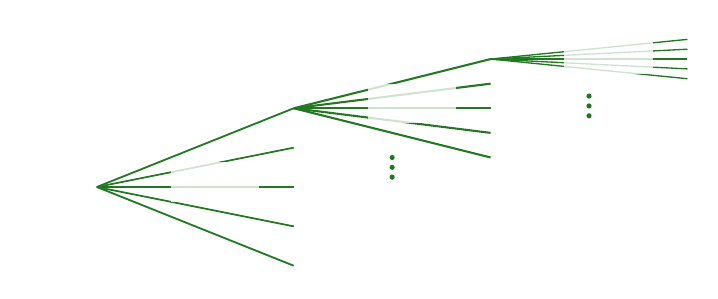 \label{fig:degradeToHanKobayashi1}
		\hspace*{-10mm}
	}
	\hfill
	\subfigure[Additionally setting $R_{11}=0$.]{
		\hspace*{-3mm}
		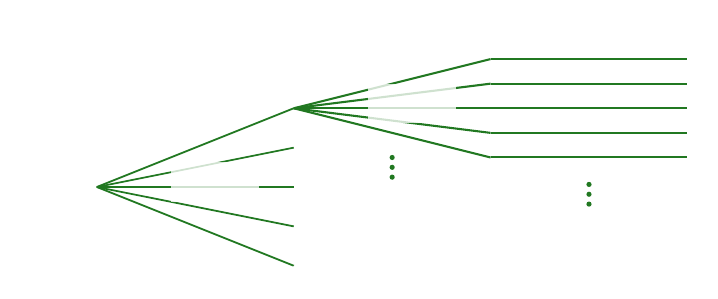 \label{fig:degradeToHanKobayashi2}
		\hspace*{-3mm}
	}
	\caption{Codebook structure for the first transmitter, when third user pair is not present.}
	\label{fig:degradeToHanKobayashi}
\end{figure}

To see this, we can simplify the coding scheme as follows. Consider the first transmitter. The Marton penalty term $I(X_{12}; X_{13} \cond U_1, Q)$ in conditions \eqref{eq:resultIneq1}, \eqref{eq:resultMaxExcess1}, and \eqref{eq:resultMaxExcess2} vanishes and we can thus choose $\tilde R_{12}=R_{12}$ and $\tilde R_{13}=R_{13}$, and furthermore, $R_{13}=0$. The codebooks then have the structure of a superposition code with three layers, namely $u_1^n$ (at rate $R_{10}$), $x_{12}^n$ (at rate $R_{12}$), and $x_1^n$ (at rate $R_{11}$), as depicted in Figure~\ref{fig:degradeToHanKobayashi1}. 

Note that there are constraints in choosing the conditional distributions underlying the superposition code, since $x_{12}$ is tied to $x_1$ by the channel definition. To implement a general superposition code with two layers as in the Han--Kobayashi scheme, let $R_{11}=0$. For each $x_{12}^n$ sequence, only a single $x_1^n$ sequence is generated, see Figure~\ref{fig:degradeToHanKobayashi2}.  Since the $x_1^n$ sequences are drawn conditioned on both $x_{12}^n$ and $u_1^n$, this is the same as generating a superposition codebook with layers $u_1^n$ (at rate $R_{10}$) and $x_1^n$ (at rate $R_{12}$).

The receivers in the general scheme perform simultaneous non-unique decoding and will thus act correctly even when the codebook structure is simplified in this fashion. The conditions in Theorem~\ref{thm:ach} simplify to the conditions in the Han--Kobayashi region.

\vspace{2mm} 
\section{Acknowledgment}
The author is grateful to  Professors Abbas El Gamal and Young-Han Kim for helpful discussions about the material presented in this paper.

\vspace*{5mm} 
\bibliographystyle{IEEEtran}
\bibliography{IEEEabrv,../unified-bibtex/references-unified,../unified-bibtex/myPublications}

\vspace*{7mm}  
\appendix
\section*{Error probability analysis}
\label{sec:errorProbAnalysis}
\newcommand{\threethreetwo}{}  

The analysis proceeds analogously to~\cite{Bandemer:2011:3DicMarton} (and more completely,~\cite{Bandemer:2011:PhdThesis}). In particular, conditions~\eqref{eq:resultIneq1} to~\eqref{eq:resultIneq2} arise from analyzing the error events related to codebook generation. As an example for an error event arising at the decoder, consider
\begin{align}
	\Er_{\threethreetwo}
	&= \bigl\{ 
	\bigl( 
		U^n_1(1),
		X^n_{12}(1, L_{12}^{(1, 1, m_{13} )}), 
		X^n_{13}(1, l_{13}),  
		X_1^n(1, L_{12}^{(1, 1, m_{13} )}, l_{13}, m_{11} ),  
		\notag  \\* 
	&\ \qquad  
		U^n_2(m_{20}), X^n_{21}(m_{20}, l_{21}),
		U^n_3(1), X^n_{31}(1, l_{31}),
		S_1^n( m_{20}, l_{21}, 1, l_{31} ), 
		Y_1^n 
	\bigr)
	\in \Typ \notag \\*
	&\ \qquad \text{for some } m_{13} \neq 1, l_{13} \notin \mathcal L_{13}(1), m_{11}, 
	m_{20} \neq 1, l_{21}, l_{31} \neq L_{31}^{(1,1,1)} \bigr\}.
	\label{eq:332_errorevent}
\end{align}
As a representative special case, we show that the condition indexed by $i=3$, $j=3$, and $k=2$ in Theorem~\ref{thm:ach} as detailed in~\eqref{eq:332_expanded} implies that the probability of this event vanishes asymptotically. In particular, we are going to show the case $j'=2, k'=1$, i.e., we prove that convergence is a consequence of the second $\min$ term in \eqref{eq:332_expanded}.
Using the union bound as in~\cite[Section 5.2.2]{Bandemer:2011:PhdThesis}, the relevant probability satisfies
\begin{align}
	&\P(\Er_{\threethreetwo} \cap \Es \cond \Er_\text{e}^c) \leq 2^{n(\tilde R_{13} + R_{11} + R_{20}+H(U_2|U_1,X_{12},U_3,Y_1)-H(U_2) )} \ P_2,
	\label{eq:332_development}
\end{align}
where $\Er_\text{e}^c$ denotes that no encoding error has occurred (ensured by conditions~\eqref{eq:resultIneq1} to~\eqref{eq:resultIneq2}), $\Es$ is defined as in~\cite{Bandemer:2011:PhdThesis}, and
\begin{align*}
	P_2 &= \P\bigl\{  
		(u_1^n,x_{12}^n,X^n_{13}(1,l_{13}), X_1^n(1,L_{12}^{(1,1,1)}, l_{13}, m_{11}), \notag \\
	& \hspace{10.5mm} 
		u_2^n, X^n_{21}(m_{20}, l_{21}),
		u^n_3, X^n_{31}(1, l_{31}), 
		y_1^n) \in \Typ \notag \\
	& \hspace{10.5mm} 
		\text{for some $l_{21}$, $l_{31} \neq L_{31}^{(1,1,1)}$}
		\condbig \Er_\text{e}^c, u_1^n, x_{12}^n, u_2^n, u_3^n, y_1^n \bigr\},
\end{align*}
where $(u_1^n, x_{12}^n, u_2^n, u_3^n, y_1^n) \in \Typ$. We bound $P_2$ as follows.
\begin{align*}
	P_2 
	&\leq \P\bigl\{  
		(u_1^n,x_{12}^n,X^n_{13}(1,l_{13}), X_1^n(1,L_{12}^{(1,1,1)}, l_{13}, m_{11}), \notag \\
	& \hspace{10.5mm} 
		u_2^n, 
		u^n_3, 
		y_1^n) \in \Typ \notag 
		\condbig \Er_\text{e}^c, u_1^n, x_{12}^n, u_2^n, u_3^n, y_1^n \bigr\},
\end{align*}
where the probability is increased by omitting parts of the typicality requirement. This step replaces the application of Corollary A.2 in~\cite{Bandemer:2011:PhdThesis} and simplifies the proof. It also leads to the slight improvement in the deterministic case as discussed above.
The following steps are fairly conventional.
\begin{align*}
	P_2 
	&\leq \hspace{-7mm} \sum_{\substack{(x_{13}^n,x_1^n)\in \Typ(X_{13},\\X_1\cond u_1^n, x_{12}^n, u_2^n, u_3^n, y_1^n)}} 
	\hspace{-7mm} \P\big\{ X_{13}^n(1,l_{13})=x_{13}^n, \\[-7mm]
	&\hspace{28mm} X_1^n(1,L_{12}^{(1,1,1)}, l_{13}, m_{11})=x_1^n \condbig \\
	&\hspace{28mm} \Er_\text{e}^c, u_1^n, x_{12}^n, u_2^n, u_3^n, y_1^n
	\big\} \\
	&\leq \hspace{-7mm} \sum_{\substack{(x_{13}^n,x_1^n)\in \Typ(X_{13},\\X_1\cond u_1^n, x_{12}^n, u_2^n, u_3^n, y_1^n)}} 
	\hspace{-7mm} \P\{ X_{13}^n(1,l_{13})=x_{13}^n \cond U_1^n(1)=u_1^n \} \\[-7mm]
	&\hspace{28mm} \P\{ X_1^n(1,L_{12}^{(1,1,1)}, l_{13}, m_{11})=x_1^n \cond \\
	&\hspace{33mm} u_1^n, x_{12}^n, x_{13}^n \} \\
	&\leq 2^{n(H(X_{13},X_1\cond U_1,X_{12},U_2,U_3,Y_1)} \\
	&\hspace{10mm} \cdot 2^{n(- H(X_{13}\cond U_1) - H(X_1\cond U_1,X_{12},X_{13}))}.
\end{align*}
Substituting back into~\eqref{eq:332_development}, we have
\begin{align*}
	\P(\Er_{\threethreetwo} \cap \Es \cond \Er_\text{e}^c)
	&\leq 2^{n(\tilde R_{13} + R_{11} + R_{20})} \\
	&\quad \cdot 2^{n(H(U_2|U_1,X_{12},U_3,Y_1)-H(U_2) )} \\
	&\quad \cdot 2^{n(H(X_{13},X_1\cond U_1,X_{12},U_2,U_3,Y_1)} \\
	&\quad \cdot 2^{n(- H(X_{13}\cond U_1) - H(X_1\cond U_1,X_{12},X_{13}))} \\
	&= 2^{n(\tilde R_{13} + R_{11} + R_{20})} \\  
	&\quad \cdot 2^{n(-I(X_{12};X_{13}\cond U_1)-H(Y_1\cond U_1,X_{12},U_3))} \\
	&\quad \cdot 2^{nH(S_1'\cond U_2,U_3)},
\end{align*}
which tends to zero as $n\to\infty$ if
\begin{align*}
	\tilde R_{13} + R_{11} + R_{20} + H(S_1'\cond U_2,U_3) &< H(Y_1\cond U_1,X_{12},U_3) + I(X_{12};X_{13}\cond U_1).
\end{align*}
Noting that 
\begin{align*}
	H(S_1'\cond U_2,U_3,S_1) &= H(S_1'\cond S_1), \\
	H(Y_1\cond U_1,X_{12},U_3,X_1,X_2,X_3) &= H(S_1'\cond S_1),
\end{align*}
and subtracting this term from both sides of the last inequality leads to 
\begin{align*}
	\tilde R_{13} + R_{11} + R_{20} + I(S_1;S_1'\cond U_2,U_3) &< I(X_1,X_2,X_3; Y_1\cond U_1,X_{12},U_3) + I(X_{12};X_{13}\cond U_1),
\end{align*}
which is the second line in~\eqref{eq:332_expanded}. 

When transitioning from~\eqref{eq:332_errorevent} to~\eqref{eq:332_development}, the union bound was applied to the indices $m_{11}$, $l_{13}$ and $m_{20}$, while the indices $l_{21}$ and $l_{31}$ where handled later by omitting terms from the typicality requirement. The latter omission is the technical reason for the saturation effects. By varying which subset of indices $\{m_{20},l_{21},l_{31}\}$ is treated by the union bound, we can obtain the remaining lines in~\eqref{eq:332_expanded}, corresponding to other modes of saturation. In all cases, the indices $m_{11}$ and $l_{13}$ are treated by the union bound, since they correspond to the intended message and thus saturation is not desirable.

\end{document}